\documentclass[aps,pre,twocolumn]{revtex4-1}
\usepackage{xcolor,hyperref}
\hypersetup{
   colorlinks,
   linkcolor={blue!50!black},%{red!80!black},
   citecolor={blue!50!black},
   urlcolor={blue!80!black}
}
\usepackage{amsmath}
\usepackage{amssymb}
\usepackage{color}
\usepackage{hyperref}
\usepackage{bm}

\usepackage{graphicx}
\usepackage{changes}

\DeclareMathAlphabet{\mathitbf}{OML}{cmm}{b}{it}

\renewcommand{\th}{^{\mbox{\tiny th}}}

\newcommand{\xv}{\mathitbf x}

\newcommand{\zerovector}{\mathBold 0}
\newcommand{\calBold}[1]{\mbox{\boldmath${\cal #1}$}}
\newcommand{\mathBold}[1]{\mbox{\boldmath$#1$}}
\newcommand{\dbar}{{\,\mathchar'26\mkern-12mu d}}

\newcommand{\sFrac}[2]{{\textstyle\frac{#1}{#2}}}

\newcommand{\taubar}{\tau\kern-0.5em\raise0.15ex\hbox{\tiny $\circ$}}

\begin{document}
\title{Mechanical properties of simple computer glasses}
\author{Edan Lerner}%${}^{1}$ and Eran Bouchbinder${}^{2}$}
\affiliation{Institute for Theoretical Physics, University of Amsterdam, Science Park 904, 1098 XH Amsterdam, The Netherlands}% \\ ${}^2$Chemical and Biological Physics Department, Weizmann Institute of Science, Rehovot 7610001, Israel}
%\affiliation{${}^1$ Institute for Theoretical Physics, University of Amsterdam, Science Park 904, 1098 XH Amsterdam, The Netherlands}% \\ ${}^2$Chemical and Biological Physics Department, Weizmann Institute of Science, Rehovot 7610001, Israel}

\begin{abstract}
Recent advances in computational glass physics enable the study of computer glasses featuring a very wide range of mechanical and kinetic stabilities. The current literature, however, lacks a comprehensive data set against which different computer glass models can be quantitatively compared on the same footing. Here we present a broad study of the mechanical properties of several popular computer glass forming models. We examine how various dimensionless numbers that characterize the glasses' elasticity and elasto-plasticity vary under different conditions --- in each model and across models --- with the aim of disentangling the model-parameter-, external-parameter- and preparation-protocol-dependencies of these observables. We expect our data set to be used as an interpretive tool in future computational studies of elasticity and elasto-plasticity of glassy solids.
\end{abstract}

\maketitle

\section{introduction}
\label{introduction}

Computational studies of glass formation and deformation constitute a substantial fraction of the research conducted in relation to these problems. The attention drawn by this line of work has been on the rise recently due to several methodological developments that allow investigators to create computer glasses with a very broad variation in the degree of their mechanical and kinetic stability. These include the ongoing optimization of Graphics-Processing-Units (GPU)-based algorithms \cite{GLASER201597,RUMD} that now offer the possibility to probe several orders of magnitude in structural relaxation rates in the supercooled liquid regime \cite{Coslovich2018}. A sampling method based on a generalized statistical ensemble has been shown to yield well-annealed states \cite{turci_prx_2017}. In groundbreaking work of Ninarello and coworkers \cite{berthier_prx}, inspired by previous advances \cite{itamar_swap}, a glass forming model was optimized such to stupendously increase the efficiency of the Swap Monte Carlo algorithm, allowing the equilibration of supercooled liquids down to unprecedented low temperatures, while remaining robust against crystallization. In \cite{fsp_arXiv} a model and algorithm were put forward that allows to create extremely stable computer glasses, albeit with a protocol which is not physical. Mechanical annealing by means of oscillatory shear was also recently shown to be an efficient protocol for creating stable glasses \cite{sri_annealing_2018}. Finally, numerical realizations of experimental vapor deposition protocols \cite{Ediger_review_2017} have shown good success in creating well-annealed glasses \cite{singh2013ultrastable, PhysRevLett.119.188002}. 

This recent proliferation of methods for creating stable computer glasses highlights the need for approaches to meaningfully and quantitatively compare between the various glasses created by these methods. In particular, it is important to disentangle the effects of parameter choices --- both in the interaction potentials that define computer glass formers, and choices of external control parameters --- from the effects of annealing near and below the models' respective computer glass transition temperatures. In addition, in some cases it is useful to quantitatively assess the effective distance a given computer glass is positioned away from the unjamming point --- the loss of rigidy seen e.g.~upon decompressing essemblies of repulsive soft spheres \cite{ohern2003,van_hecke_review,liu_review}.% --- affects elastic and elasto-plastic responses. 

This work is aimed towards establishing how elastic properties and elasto-plastic responses of simple computer glasses depend on various key external and internal control parameters, how they change between different models, and how they are affected by the preparation protocol of glasses. In order to disentangle annealing effects from model- and external-parameter dependences, we exploit the observation that creating computer glasses by instantaneous quenches of high energy states to zero temperature defines an ensemble of configurations whose elastic properties can be meaningfully and quantitatively compared between models and across different parameter regimes. 

%%%%%%%%%%%%%%%%%%%%%%%%%%%%%%%%%%%%%%%%%%%%%
\begin{figure}[!ht]
\centering
\includegraphics[width = 0.50\textwidth]{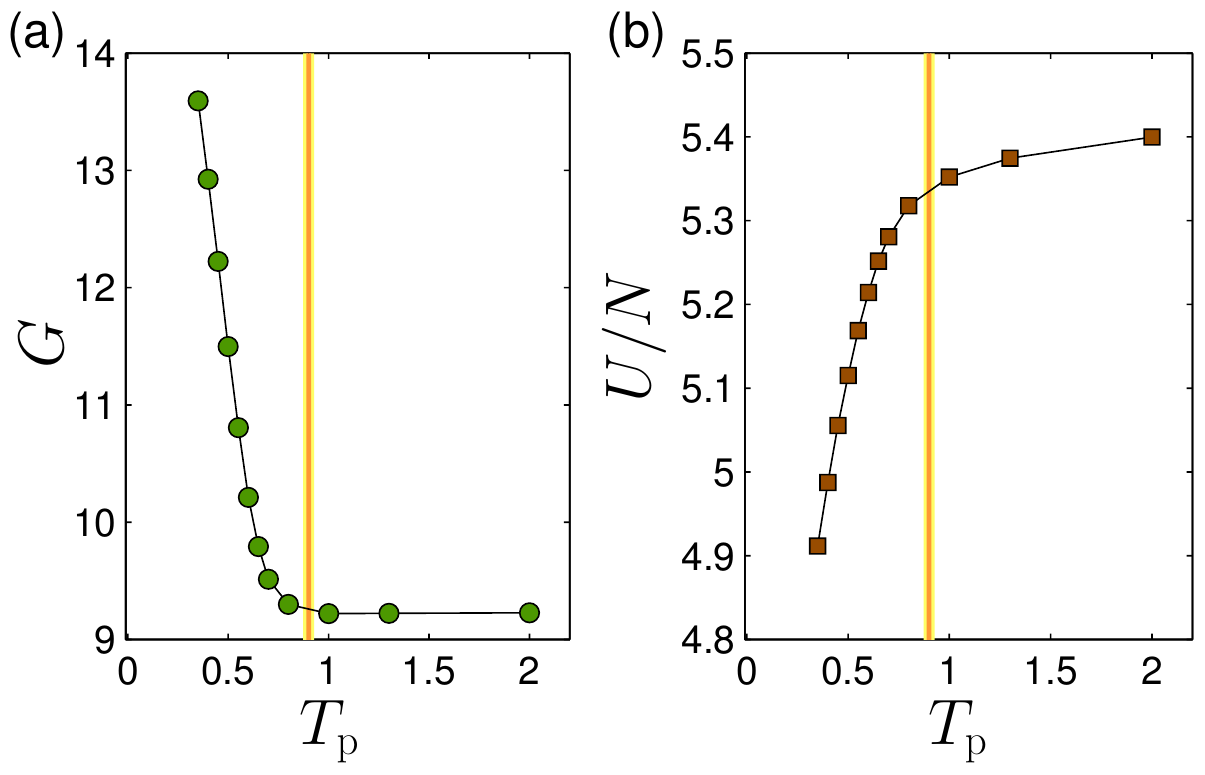}
\caption{\footnotesize (a) Sample-to-sample mean athermal shear modulus $G$ measured in inherent states that underlie liquid states at equilibrium parent temperatures $T_{\rm p}$ of the POLY model, see Sect.~\ref{swap} for model details. The vertical line approximates the crossover temperature above which several elastic properties saturate. (b) Sample-to-sample mean inherent state potential energy-per-particle $U/N$. Interestingly, while $G$ saturates above the crossover temperature, $U/N$ does not. In this work we focus on several observables that feature a saturation as seen for $G$ in panel (a).}
\label{plateau_fig}
\end{figure}
%%%%%%%%%%%%%%%%%%%%%%%%%%%%%%%%%%%%%%%%%%%%%%%%%%%%%%

The existence of the aformentioned ensemble is demonstrated in Fig.~\ref{plateau_fig}, where we plot measurements of the sample-to-sample mean athermal shear modulus (see definition below) of underlying inherent states of parent equilibrium configurations (labelled by their equilibrium temperature $T_{\rm p}$) of a simple glass-forming model (see details in Sect.~\ref{swap} below). This high-temperature saturation of elastic properties of very poorly annealed glassy states appears to be a generic feature of computer glasses \cite{Sastry1998,SASTRY2002267,Ashwin2004,cge_paper,LB_modes_2019}. We therefore carry out in what follows a comparative study of elastic properties of different computer glass models created by instantaneous quenches from high energy states. Our analyses of elastic properties of instantanously-quenched glasses are compared against the behavior of the same key observables measured in a variant of the glass forming model introduced in \cite{berthier_prx} that can be annealed very deeply below the conventional computer glass transition temperature. This allows us to compare the relative protocol- and parameter-induced variation in these key observables. 

In the same spirit, we also investigate the elasto-plastic steady state as seen by deforming our instantaneously-quenched glasses using an athermal, quasistatic shear protocol, that gets rid of any rate effects associated with finite deformation rate  and finite temperature protocols.  We anticipate our results to constitute a benchmark for quantitative assessment of other glasses in future studies of elasticity, elasto-plasticity and glass formation. 

This paper is structured as follows; in Sect.~\ref{models} we spell out the models employed in our study, and list the physical observables that were calculated in those models. Sect.~\ref{results} presents various data sets that characterize the elasticity and elasto-plasticity of the computer glasses we have investigated, and discusses various points of interest and connections to related previous work. Our work is summarized in Sect.~\ref{summary}. 

\section{Models, methods and observables}
\label{models}

In this Section we provide details about the model glass formers we employed in our study, and explain the methods used to create glassy samples. We then spell out the definitions of all reported observables. 

\subsection{Computer glasses}

We have studied 4 model glass formers in $\dbar\!=\!3$ dimenions. We have created ensembles of at least 1000 configurations of at least $N\!=\!8000$ particles for each model system, and for each value of the respective control parameter (see below). 
% For the first three models described below (Hertzian, Kob-Andersen binary Lennard-Jones and inverse-power-law) glasses were created by first placing the particles randomly on a cubic lattice, followed by minimizing the potential energy. For the Hertzian system we employed the FIRE minimization algorithm, whereas for the  by means of a conventional nonlinear conjugate gradient algorithm. 

\subsubsection{Inverse-power-law}
\label{ipl_model}
The inverse-power-law (IPL) model is a 50:50 binary mixture of `large' and `small' particles of equal mass $m$. Pairs of particles $i,j$ at distance $r_{ij}$ from each other interact via the inverse-power law pairwise potential
\begin{equation}\label{ipl_pairwise_potential}
\varphi_{\mbox{\tiny IPL}}(r_{ij}) = \varepsilon\left( \sFrac{\lambda_{ij}}{r_{ij}} \right)^\beta\,,
\end{equation}
where $\varepsilon$ is a microscopic energy scale. Distances in this model are measured in terms of the interaction lengthscale $\lambda$ between two `small' particles, and the rest are chosen to be $\lambda_{ij}\!=\!1.18\lambda$ for one `small' and one `large' particle, and $\lambda_{ij}\!=\!1.4\lambda$ for two `large' particles. In finite systems under periodic boundary conditions, a variant of the IPL model with a finite interaction range should be employed, otherwise the potential is discontinuous due to the periodic boundary conditions. We chose the form
\begin{equation}\label{ipl_pairwise_potential}
\varphi_{\mbox{\tiny IPL}}(r_{ij}) = \left\{ \begin{array}{ccc}\varepsilon\left[ \left( \sFrac{\lambda_{ij}}{r_{ij}} \right)^\beta + \sum\limits_{\ell=0}^q c_{2\ell}\left(\sFrac{r_{ij}}{\lambda_{ij}}\right)^{2\ell}\right]&,&\sFrac{r_{ij}}{\lambda_{ij}}\le x_c\\0&,&\sFrac{r_{ij}}{\lambda_{ij}}> x_c\end{array} \right.\!\!,
\end{equation}
where $x_c$ is the dimensionless distance for which $\varphi_{\mbox{\tiny IPL}}$ vanishes continuously up to $q$ derivatives. The coefficients $c_{2\ell}$, determined by demanding that $\varphi$ vanishes continuously up to $q$ derivatives, are given by
\begin{equation}
c_{2\ell} = \frac{(-1)^{\ell+1}}{(2q\!-\!2\ell)!!(2\ell)!!}\frac{(\beta\!+\!2q)!!}{(\beta\!-\!2)!!(\beta\!+\!2\ell)}x_c^{-(\beta+2\ell)}\,.
\end{equation}
For $\beta\!<\!16$ we chose the largest $x_c$ possible, which is $L/2\!\times\!(1.4\lambda)^{-1}$ with $L$ denoting the linear size of the system. This cutoff, set to be exactly half the system's length for the `large'-`large' pair interactions, results in a model that is the closest we can approach the full-blown IPL potential energy in which \emph{all} pairs of particles interact, with no interaction cutoff. For $\beta\!\ge\!16$ the cutoff no longer plays a role, then we chose $x_c\!=\!1.5$, and $\hat{\rho}\!\equiv\!N/V\!=\!0.82$ for computational efficiency, and $q\!=\!2$ for all simulations ($V\!=\! L^3$ is the volume). The effects of varying the dimensionless cutoff $x_c$ and system size $N$ are discussed in Appendix~\ref{ipl_appendix}. 

The control parameter of interest for this system is the exponent $\beta$ of the inverse-power-law pairwise interaction, which we varied between $\beta\!=\!4$ and $\beta\!=\!256$. Glassy samples were created by placing $N\!=\!8000$ for $\beta\!>\!6$, $N\!=\!16000$ for $\beta=6$, or $N=32000$ for $\beta\!=\!4$ (see Appendix~\ref{ipl_appendix} for discussion) particles randomly on a cubic lattice and minimizing the potential energy by a conjugate gradient minimization.

\subsubsection{Hertzian spheres}
\label{HRTZ_model}
The Hertzian spheres model (HRTZ) we employ is a 50:50 binary mixture of soft spheres with equal mass $m$ and a 1:1.4 ratio of the radii of small and large particles. The units of length $\lambda$ are chosen to be the diameter of the small particles, and $\varepsilon$ denotes the microscopic units of energy. Pairs of particles whose pairwise distance $r_{ij}$ is smaller than the sum of their radii $R_i\!+\!R_j$ interact via the Hertzian pairwise potential
\begin{equation}
\varphi_{\mbox{\tiny Hertz}}(r_{ij},R_i,R_j) = \sFrac{2\varepsilon}{5\lambda^{5/2}}\big( (R_i + R_j) - r_{ij} \big)^{5/2}\,,
\end{equation}
and $\varphi_{\mbox{\tiny Hertz}}\!=\!0$ otherwise. 

In this model we control the imposed pressure; glassy samples at target pressures of $p\!=\!10^{-1},10^{-2},10^{-3},10^{-4},10^{-5}$ were created by combining a Berendsen barostat \cite{berendsen} into the FIRE minimization algorithm \cite{fire}. Initial states at the highest pressure were created by placing particles randomly on a cubic lattice, followed by minimizing the potential energy. Subsequent lower pressure glasses were created by changing the target pressure and relaunching the minimization algorithm.

\subsubsection{Kob-Andersen binary Lennard-Jones}
We employ a slightly modified variant of the well-studied Kon-Andersen binary Lennard-Jones (KABLJ) glass former \cite{kablj}, which is perhaps the most widely studied computer glass model. Our variant of the KABLJ model is a binary mixture of 80\% type A particles and 20\% type B particles, that interact via the pairwise potential
\begin{eqnarray}
\varphi_{\mbox{\tiny KABLJ}}(r_{ij})& = &4\varepsilon_{ij}\left( \big(\sFrac{\lambda_{ij}}{r_{ij}}\big)^{12} - \big(\sFrac{\lambda_{ij}}{r_{ij}}\big)^{6} \right. \nonumber \\
& & \left.\ \ \ \ \ \ \ \  + \ c_4\big(\sFrac{r_{ij}}{\lambda_{ij}}\big)^4 + c_2\big(\sFrac{r_{ij}}{\lambda_{ij}}\big)^2 + c_0\right),
\end{eqnarray}
if $r_{ij}/\lambda_{ij}\!\le\!2.5$, and $\varphi_{\mbox{\tiny KABLJ}}\!=\!0$ otherwise. Lengths are expressed in terms of $\lambda_{AA}$, then $\lambda_{AB}\!=\!4/5$ and $\lambda_{BB}\!=\!22/25$. Energies are expressed in terms of $\varepsilon_{AA}$, then $\varepsilon_{AB}\!=\!3/2$ and $\varepsilon_{BB}\!=\!1/2$. Both particle species share the same mass $m$. The coefficients $c_4,c_2$ and $c_0$ are chosen such that $\varphi_{\mbox{\tiny KABLJ}}$, $\varphi_{\mbox{\tiny KABLJ}}'$ and $\varphi_{\mbox{\tiny KABLJ}}''$ vanish at $r_{ij}/\lambda_{ij}\!=\!5/2$. In this model we control the density $\rho\!\equiv\! N/V$ with $V$ denoting the volume.

\subsubsection{Polydisperse soft spheres}
\label{swap}
The computer glass model we employed is a slightly modified variant of the model put forward in \cite{berthier_prx}. We enclose $N$ particles of equal mass $m$ in a square box of volume $V\!=\! L^3$ with periodic boundary conditions, and associate a size parameter $\lambda_i$ to each particle, drawn from a distribution $p(\lambda)\!\sim\!\lambda^{-3}$. We only allow $\lambda_{\mbox{\tiny min}}\!\le\!\lambda_i\le\!\lambda_{\mbox{\tiny max}}$ with $\lambdabar\!\equiv\!\lambda_{\mbox{\tiny min}}$ forming our units of length, and $\lambda_{\mbox{\tiny max}}\!=\!2.22\lambdabar$. The number density $N/V\!=\!0.58\lambdabar^{-3}$ is kept fixed. Pairs of particles interact via the same pairwise interaction give by Eq.~(\ref{ipl_pairwise_potential}). We chose the parameters $x_c\!=\!1.4, n\!=\!10$, and $q\!=\!3$. The pairwise length parameters $\lambda_{ij}$ are given by 
\begin{equation}
\lambda_{ij} = \sFrac{1}{2}(\lambda_i + \lambda_j)(1 - n_a|\lambda_i - \lambda_j|)\,.
\end{equation}
Following \cite{berthier_prx} we set the non-additivity parameter $n_a\!=\!0.1$. In what follows energy is expressed in terms of $\varepsilon$, temperature is expressed in terms of $\varepsilon/k_B$ with $k_B$ the Boltzmann constant, stress, pressure, and elastic moduli are expressed in terms of $\varepsilon/\lambdabar^3$. This model is referred to in what follows as POLY. 

Ensembles of equilibrium states of the POLY model were created using the Swap Monte Carlo method \cite{itamar_swap,berthier_prx}; within this method, trial moves include exchanging (swapping) the size parameters of pairs of particles, in addition to the conventional random displacements of particles. For each temperature we have simulated 50 independent systems of $N\!=\!8000$ particles, and collected 20 configurations for each system that were separated by at least the structural relaxation time (here time is understood as Monte-Carlo steps) as measured by the stress autocorrelation function, resulting in equilibrium ensembles of 1000 members for each parent temperature $T_{\rm p}$. Ensembles of inherent states were created by performing an instantaneous quench of equilibrium states from each parent temperature by means of a conjugate gradient minimization of the potetial energy. 

We note that since particle size parameters are sampled from a rather broad distribution, and our simulated systems are of only $N\!=\!8000$ particles, very large finite-size sampling-induced fluctuations of the equilibrium energy of different systems (which are entirely absent in e.g.~binary systems such as the KABLJ) can occur; a description of how we reduced these fluctuations --- which can affect various fluctuation measures described in what follows --- is provided in Appendix~\ref{fluctuations}.

\subsection{Observables}
\label{observables}

In what follows we will denote by $\xv_i$ the 3-dimensional coordinate vector of the $i\th$ particle, then $\xv_{ij}\!\equiv\!\xv_j\!-\!\xv_i$ is the vector distance between the $i\th$ and $j\th$ particles, and $r_{ij}\!\equiv\!\sqrt{\xv_{ij}\!\cdot\!\xv_{ij}}$ is the pairwise distance between them. We also omit the explicit mentioning of dimensional observables' units, for the sake of simplifying our notations; those observables should be understood as expressed in the appropriate microscopic units.

In all computer glasses considered in this work, pairs of particles $i,j$ interact via a radially-symmetric pairwise potential $\varphi_{ij}\!=\!\varphi_{ij}(r_{ij})$, then the potential energy reads
\begin{equation}
U = \sum_{i<j}\varphi_{ij}\,.
\end{equation}
The (simple shear) stress in athermal glasses is given by
\begin{equation}
\sigma = \frac{1}{V}\frac{\partial U}{\partial\gamma}\,.
\end{equation}
We also consider the shear and bulk moduli, defined as
\begin{equation}\label{shear_modulus}
G = \frac{\frac{\partial^2U}{\partial\gamma^2} - \frac{\partial^2U}{\partial\gamma\partial\xv}\cdot{\calBold{M}}^{-1}\cdot\frac{\partial^2U}{\partial\xv\partial\gamma}}{V}\,,
\end{equation}
and
\begin{equation}\label{bulk_modulus_def}
K = \frac{\frac{\partial^2U}{\partial\eta^2} - \frac{\partial^2U}{\partial\eta\partial\xv}\cdot{\calBold{M}}^{-1}\cdot\frac{\partial^2U}{\partial\xv\partial\eta}}{\dbar^2V} + p\,,
\end{equation}
respectively, where the pressure $p$ is given by
\[
p = -\frac{1}{V\dbar}\frac{\partial U}{\partial\eta}\,,\quad \calBold{M}\equiv\frac{\partial^2U}{\partial\xv\partial\xv}
\]
is the Hessian matrix, and $\eta,\gamma$ parametrize the strain tensor
\begin{equation}\label{strain_tensor}
\calBold{\epsilon} = \frac{1}{2}\left( \begin{array}{ccc}2\eta + \eta^2& \gamma  + \gamma\eta& 0 \\ \gamma  + \gamma\eta & 2\eta + \eta^2 + \gamma^2&0 \\ 0 & 0 &2\eta + \eta^2 \end{array}\right)\,.
\end{equation}
To quantify the effect of nonaffinity on the bulk modulus $K$, we also consider the nonaffine term alone (cf.~Eq.~(\ref{bulk_modulus_def})), namely
\begin{equation}\label{nonaffine_term}
K_{\mbox{\tiny na}}\equiv \frac{1}{\dbar^2V}\frac{\partial^2U}{\partial\eta\partial\xv}\cdot{\calBold{M}}^{-1}\cdot\frac{\partial^2U}{\partial\xv\partial\eta}\,.
\end{equation}

The Poisson's ratio is given by 
\begin{equation}\label{Poissons_ratio_def}
\nu \equiv \frac{3K - 2G}{6K + 2G} = \frac{3-2G/K}{6+2G/K}\,.
\end{equation}

For every model studied in what follows, we also consider an ``unstressed" potential energy
\begin{equation}\label{mock_U}
{\cal U} = \sFrac{1}{2} \sum_{i<j} \varphi_{ij}''(r_{ij}-r_{ij}^{(0)})^2\,,
\end{equation}
where $\varphi_{ij}''$ is the second derivative of the $i,j$ interaction of the original potential, and $r_{ij}^{(0)}$ is the distance between the $i\th$ and $j\th$ particles in the mechanical equilibrium state $\partial U/\partial\xv\!=\!\zerovector$ of the original potential. The potential ${\cal U}$ can be understood as obtained by replacing the original interactions by Hookean springs whose stiffnesses are inherented from the original interaction potential, and that reside exactly at their rest lengths $r_{ij}^{(0)}$ so that the springs exert no forces on the particles. The observable we focus on is then the shear modulus ${\cal G}\!\equiv\!V^{-1}\frac{d^2{\cal U}}{d\gamma^2}$ of the unstressed potential. 

%%%%%%%%%%%%%%%%%%%%%%%%%%%%%%%%%%%%%%%%%%%%%
\begin{figure*}%[!ht]
\centering
\includegraphics[width = 1.00\textwidth]{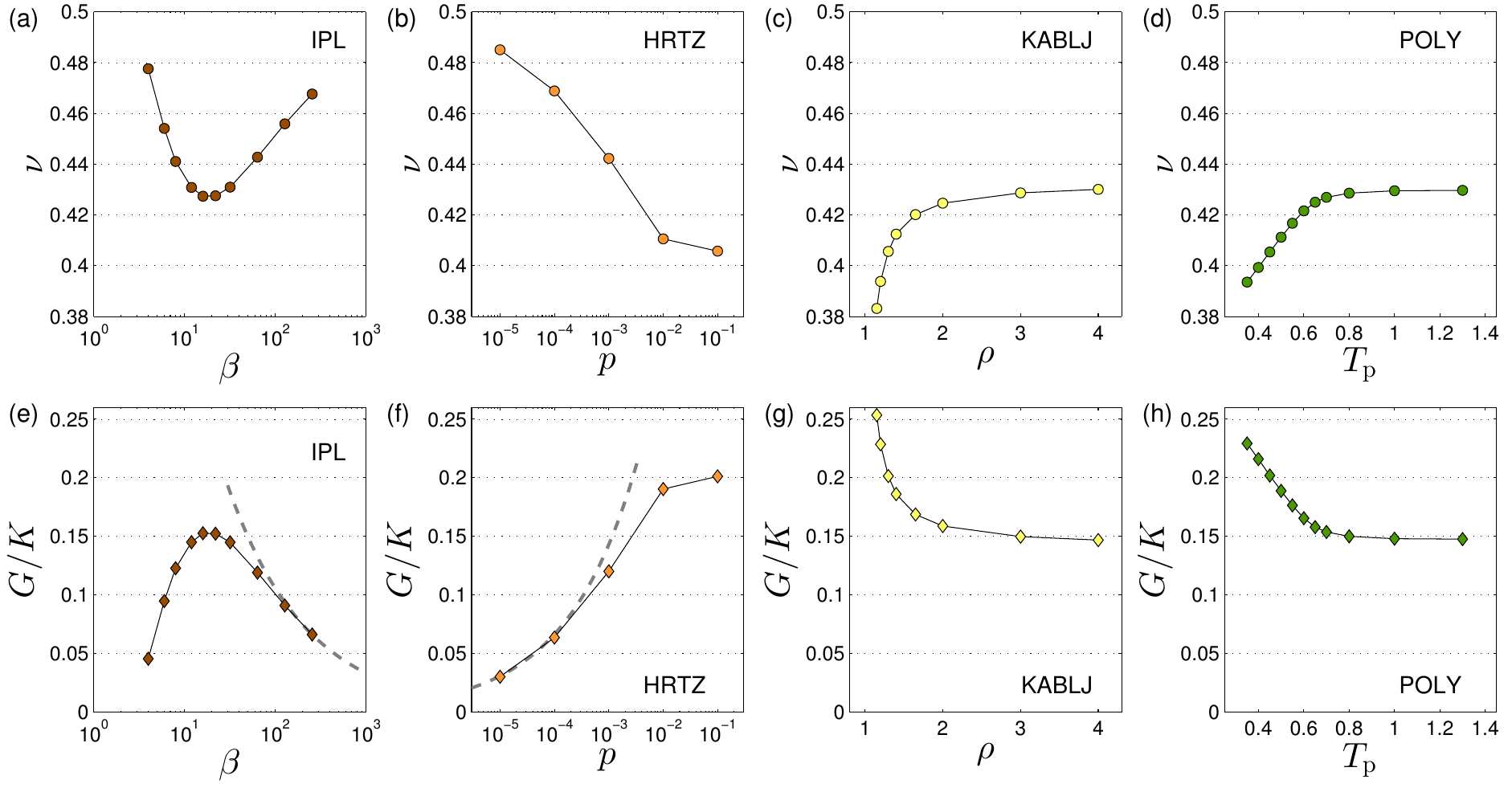}
\caption{\footnotesize (a)-(d) Sample-to-sample mean Poisson's ratio measured in our different models and ensembles of glassy solids. Here and in the following figures, panel (a) shows data for the IPL model, panel (b) shows data for the HRTZ model, panel (c) shows data for the KABLJ model, and panel (d) shows data for the POLY model. The bottom row shows the shear to bulk moduli ratio $G/K$. The dashed lines in panels (e) and (f) represent the scaling laws expected upon approaching the unjamming point.}
\label{poissons_ratio_fig}
\end{figure*}
%%%%%%%%%%%%%%%%%%%%%%%%%%%%%%%%%%%%%%%%%%%%%%%%%%%%%%

\section{Results}
\label{results}

Here we present the various data sets of dimensionless observables that describe the mechanical properties of the glasses of different models and control parameters discussed in the previous Section. 

\subsection{Poisson's ratio}
\label{poissons_ratio}
We begin with presenting data for the Poisson's ratio $\nu$ (defined in Eq.~(\ref{Poissons_ratio_def})), which is a conventional dimensionless characterizer of the elastic properties of solids \cite{Greaves2011,Saxena}, whether glassy \cite{typical_poissons_ratio_metallic_glass} or crystalline \cite{Baughman1998}. It has~also been shown to feature some correlation with the degree of ductility or brittleness of material failure \cite{Greer_2005}. Fig.~\ref{poissons_ratio_fig}(a)-(d) shows the sample-to-sample means of the Poisson's ratio measured in our ensembles of the model glasses studied. To gain insight on the behavior of $\nu$, we also plot in panels (e)-(h) the ratio $G/K$ (cf.~Eq.~(\ref{Poissons_ratio_def})) of the sample-to-sample means of the shear and bulk moduli.

Fig.~\ref{poissons_ratio_fig}a shows the Poisson's ratio of the IPL model; we observe an interesting non-monotonic behavior of $\nu$ as a function of the exponent $\beta$ that characterizes the pairwise interaction. A corresponding non-monotonic behavior of the ratio $G/K$ is also observed (Fig.~\ref{poissons_ratio_fig}e); the decrease of $G/K$ at large $\beta$ is expected: in previous work \cite{stefanz_pre} it was shown that increasing $\beta$ is akin to approaching the unjamming point of repulsive soft spheres \cite{ohern2003,liu_review,van_hecke_review}. In \cite{stefanz_pre} it was shown that $G/K$ is expected to vanish as $1/\sqrt{\beta}$, represented in Fig.~\ref{poissons_ratio_fig}e by the dashed line. 

The decrease of the ratio $G/K$ at small $\beta$ seen in Fig.~\ref{poissons_ratio_fig}e, that leads in turn to an increase in the Poisson's ratio $\nu$ for small $\beta$, is however unexpected. How can this nonmonotonicity of $G/K$ be understood?

To reveal the origin of the sharp decrease of $G/K$ at small $\beta$, we point out that following Eq.~(\ref{strain_tensor})
\begin{equation}
\frac{\partial^2U}{\partial \gamma^2} = \frac{\partial^2U}{\partial\epsilon_{xy}^2} + \frac{\partial U}{\partial \epsilon_{yy}}\,,
\end{equation}
where the second term stems from the $\gamma^2$ term in $\epsilon_{yy}$, see Eq.~(\ref{strain_tensor}). Noticing that $V^{-1}\partial U/\partial \epsilon_{yy}\!\simeq\!-p$, and following Eq.~(\ref{shear_modulus}), $G$ can be decomposed into three terms as
\begin{equation}\label{decomposition}
G = \frac{1}{V}\frac{\partial^2U}{\partial\epsilon_{xy}^2} - p  - G_{\mbox{\tiny na}}\,,
\end{equation}
where $p$ is strictly positive due to the purely-repulsive pairwise interactions of the IPL model, and $G_{\mbox{\tiny na}}\!\equiv\!\frac{1}{V}\frac{\partial^2U}{\partial\epsilon_{xy}\partial\xv}\!\cdot\!\calBold{M}^{-1}\!\cdot\!\frac{\partial^2U}{\partial\xv\partial\epsilon_{xy}}$ is strictly positive due to the positive-definiteness of $\calBold{M}$. 

%%%%%%%%%%%%%%%%%%%%%%%%%%%%%%%%%%%%%%%%%%%%%
\begin{figure}[!ht]
\centering
\includegraphics[width = 0.50\textwidth]{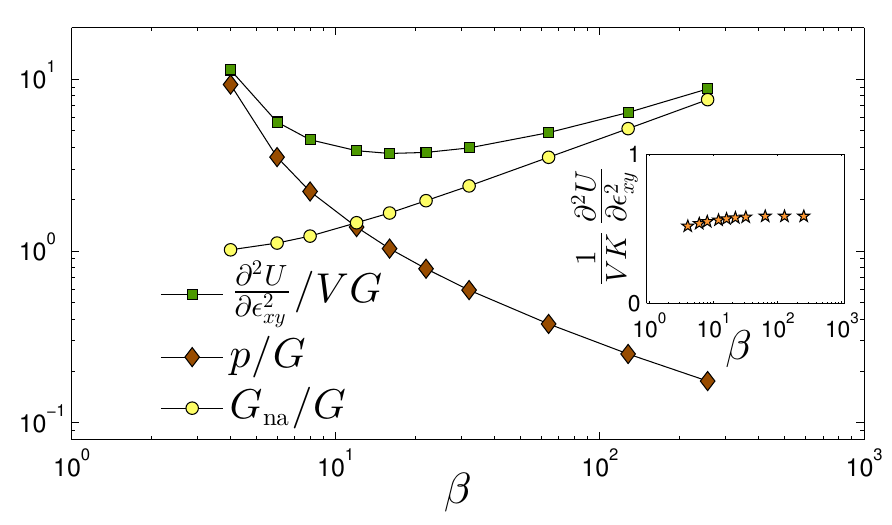}
\caption{\footnotesize Decomposition of the shear modulus into its different relative contributions as spelled out in Eq.~(\ref{decomposition}), i.e.~$G\!=\square - \Diamond - \circ$. The inset demonstrates that $V^{-1}\partial^2U/\partial\epsilon_{xy}^2\!\sim\! K$ over the entire investigated $\beta$ range. }
\label{G_contributions}
\end{figure}
%%%%%%%%%%%%%%%%%%%%%%%%%%%%%%%%%%%%%%%%%%%%%%%%%%%%%%

In Fig.~\ref{G_contributions} we show the three relative contributions to the shear modulus as explained above, as a function of the power $\beta$. Interestingly, at small $\beta$ the shear modulus is given by a near cancellation of numbers that are larger than their difference by more than an order of magnitude, similarly to the phenomenology close to the unjamming transition \cite{ohern2003,van_hecke_review,liu_review}, and, in our case, also seen at large $\beta$. However, as opposed to near unjamming where $G_{\mbox{\tiny na}}$ is responsible for the smallness of $G$, at small $\beta$ it is the contribution due to the pressure (stemming from the $\gamma^2$ term in $\epsilon_{yy}$, see Eq.~(\ref{strain_tensor})) which cancels the affine shear stiffness term $V^{-1}\partial^2U/\partial\epsilon_{xy}^2$ to produce a small $G$ compared to $K$. In the inset of Fig.~\ref{G_contributions} we show that the affine shear stiffness term scales with the bulk modulus over the entire range of $\beta$ measured, establishing that indeed the reduction of $G$ due to the pressure is responsible for decreasing $G/K$ at small $\beta$. 

% The importance of higher coordination shells for small $\beta$ can be appreciated by considering the density dependence of the Poisson's ratio $\nu$ and the ratio $G/K$ for the IPL model with a finite interaction cutoff length (see Sect.~\ref{ipl_model}), shown in Fig.~\ref{cutoff_effect_fig}. We note that the inverse-power-law interactions of the IPL model should imply the invariance of any dimensionless numbers to changes of the density \cite{Dyre_2016}. However, upon introducing a cutoff (for computational efficiency) in the pairwise interactions, the invariance becomes only approximate. In particular, for $\beta\!<\!12$ a measurable dependence of $G/K$ (and therefore also of $\nu$) is seen over a broad range of densities, indicating the greater importance of higher coordination shells for small $\beta$ in the IPL model. 

%data set shows a very large difference with the value measured with no cutoff in the interactions: we find $G/K\!\approx0.07$ for a relatively high density of $\rho\!=\!2.0$, while the $\rho\!\to\!\infty$ value (marked by the horizontal gray lines in Fig.~\ref{cutoff_effect_fig}) of $\approx\!0.04$. 

Fig.~\ref{poissons_ratio_fig}b shows the Poisson's ratio $\nu$ measured in the HRTZ system, plotted against the imposed pressure $p$. As $p\!\to\!0$ it appears that the incompressible limit $\nu\!=\!1/2$ is approached. As expected, this is a consequence of the aformentioned vanishing of the ratio $G/K$ upon approaching the unjamming point, as indeed seen in Fig.~\ref{poissons_ratio_fig}f. It is known \cite{ohern2003} that in the HRTZ model $G\!\sim\! p^{2/3}$ and $K\!\sim\! p^{1/3}$, and so one expects $G/K \!\sim\! p^{1/3}$, represented by the dashed line in Fig.\ref{poissons_ratio_fig}f. Interestingly, in both the HRTZ model and in the IPL model it appears that the onset of the scaling regime takes place at $G/K\!\approx\!0.1$.  

%%%%%%%%%%%%%%%%%%%%%%%%%%%%%%%%%%%%%%%%%%%%%
\begin{figure}[!ht]
\centering
\includegraphics[width = 0.50\textwidth]{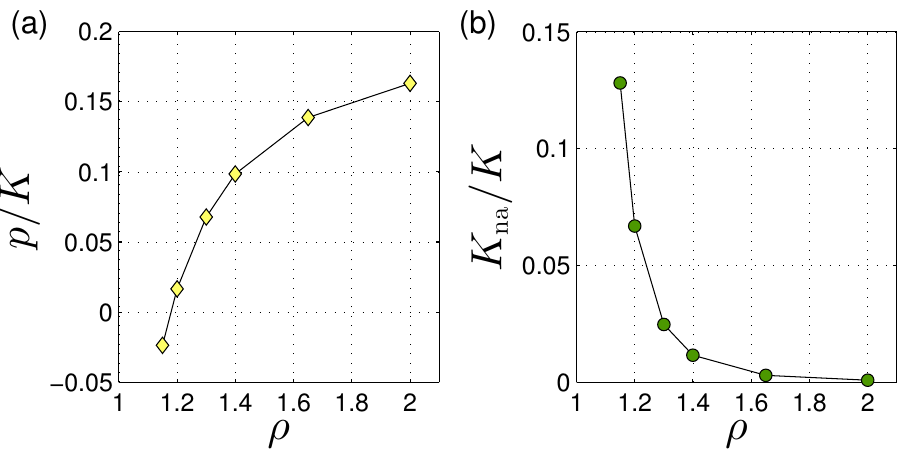}
\caption{\footnotesize (a) Pressure to bulk modulus ratio vs.~the density, for the KABLJ system. (b) The relative fraction of the nonaffine term of the bulk modulus, see text for definitions and discussion.}
\label{pressure_and_nonaffine_K_fig}
\end{figure}
%%%%%%%%%%%%%%%%%%%%%%%%%%%%%%%%%%%%%%%%%%%%%%%%%%%%%%

%%%%%%%%%%%%%%%%%%%%%%%%%%%%%%%%%%%%%%%%%%%%%
\begin{figure*}[!ht]
\centering
\includegraphics[width = 1.00\textwidth]{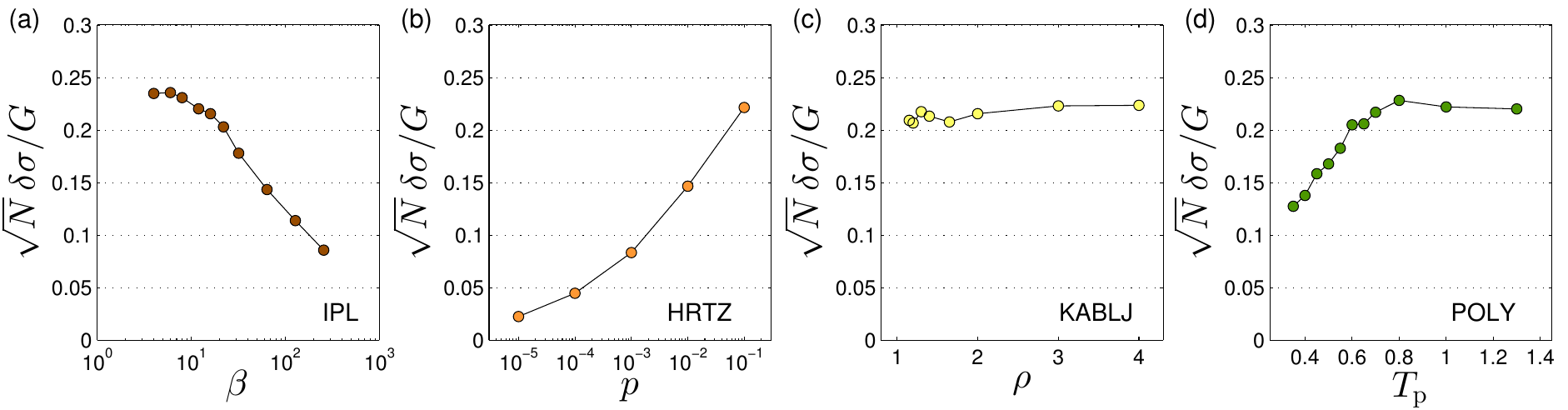}
\caption{\footnotesize Sample-to-sample standard deviations of the shear stress $\delta\sigma$, scaled by $\sqrt{N}/G$ with $G$ the mean shear modulus.}
\label{stress_flucs_fig}
\end{figure*}
%%%%%%%%%%%%%%%%%%%%%%%%%%%%%%%%%%%%%%%%%%%%%%%%%%%%%%

Fig.~\ref{poissons_ratio_fig}c shows the Poisson's ratio $\nu$ measured in the KABLJ system, plotted against the density $\rho$. Here we see that the large density $\nu$ agrees with the IPL results for $\beta\!\approx\!12$; indeed one expects the repulsive part of the KABLJ pairwise potential to dominate the mechanics at high densities \cite{jeppe}. At lower densities, the attractive part of the pairwise interactions of the KABLJ model start to play an increasingly important roll, leading to a plummet of $\nu$ as the density approaches unity. This sharp decrease is echoed by a sharp increase in $G/K$ seen in Fig.~\ref{poissons_ratio_fig}g.

To better understand these observations in the KABLJ data, we plot in Fig.~\ref{pressure_and_nonaffine_K_fig}a the ratio of the pressure to bulk modulus of the KABLJ systems, vs.~the density. As expected, the pressure decreases with decreasing density, and appears to vanish a bit below $\rho\!=\!1.2$ \cite{sri_prl_2000}.  Accompanying the vanishing of pressure is a substantial increase in nonaffine nature of displacements under compressive strains, which we quantify via the nonaffine contribution to the bulk modulus $K_{\mbox{\tiny na}}$ defined in Eq.~(\ref{nonaffine_term}). Fig.~\ref{pressure_and_nonaffine_K_fig}b shows that the relative fraction that $K_{\mbox{\tiny na}}$ amounts to in the bulk modulus grows from nearly zero at $\rho\!\ge\!2.0$ to about 13\% at $\rho\!=\!1.15$. This increase in the nonaffine contribution to the moduli, together with the contribution of the negative pressure (cf.~Eq.~(\ref{bulk_modulus_def})), can explain most of the increase of $G/K$, and the corresponding decrease of the Poisson's ratio at low densities, in the KABLJ model.

Finally, in Fig.~\ref{poissons_ratio_fig}d we show the Poisson's ratio measured in the POLY system, plotted against the equilibrium parent temperature $T_{\rm p}$ from which the ensembles of glasses were quenched. The annealing at the lowest temperature leads to a decrease of slightly more than 8\% in $\nu$. In terms of the ratio $G/K$, we observe an annealing-induced increase of over $55\%$ above the high-$T_{\rm p}$ plateau. For comparison, in \cite{CHENG20093253} an increase of nearly 20\% in $G/K$ was observed by varying the quench rate of a model of a $\mbox{Cu}_{64}\mbox{Zr}_{36}$ metallic glass over two orders of magnitude, with an associated increase of $\approx\!3.5\%$ in the Poisson's ratio, whose typical values were found around $\nu\!=\!0.41$.

We note that typical values for the Poisson's ratio of metallic glasses ranges between 0.3-0.4 \cite{typical_poissons_ratio_metallic_glass,CHENG20093253,Greer_2005}, i.e.~mostly \emph{lower} than what we observe in our simple models, with the exception of the KABLJ model, discussed in length above. We attribute the higher values of $\nu$ seen in our models that feature inverse-power-law pairwise interactions (i.e.~the IPL model, and the KABLJ at high densities) to the relative smallness of the nonaffine term in the bulk modulus. This relative smallness results in relatively larger bulk moduli (compared to shear moduli), and, in turn, to higher Poisson's ratios. Laboratory glasses experience a significant degree of annealing upon preparation, which would further \emph{reduce} their Poisson's ratio, as suggested by our measurements of the POLY system shown in Fig.~\ref{poissons_ratio_fig}h.

% In addition to the aformentioned effect, the proximity to the unjamming point that is expected in the large $\beta$ limit and the small pressure limit in the IPL and the HRTZ models, respectively, also lead to higher values of the Poisson's ratio. This increase is a result of the descrease of $G/K$ due to the faster decrease of $G$ compared to $K$ as unjamming is approached. 

\subsection{Degree of internal stresses}
\label{internal_stress}
One of the hallmark features of glasses is their structural frustration. How can the degree of structural frustration of different computer glasses be compared? Here we offer to quantitatively compare different simple computer glasses via the following observable: consider a glassy sample that is comprised of $N$ particles; consider next replacing the fixed-shape box in which the glass is confined by a box that can undergo simple shear deformation, and consider fixing the imposed shear stress (instead of the box shape) at zero. Under these conditions, the internal residual stresses of the glass would lead to some shear deformation $\delta\gamma$ of the box, that can be estimated as $\delta\gamma\!\approx\!\sigma/G$, where $\sigma$ is the as-cast shear stress of the original glass. Since $\delta\gamma$ decays with system size $N$ as $1/\sqrt{N}$ (it is $N^{-1}$ times a sum of ${\cal O}(N)$ random contributions, see Appendix~\ref{N_scaling_appendix} for numerical validation), we thus form a dimensionless characterization of glassy structural frustration by 
\begin{equation}
\delta \tilde{\sigma} \equiv \sqrt{N}\delta\gamma = \sqrt{N}\delta\sigma/G\,,
\end{equation}
where $\delta\sigma$ denotes the sample-to-sample standard deviation of the residual stresses. 

In Fig.~\ref{stress_flucs_fig} we show $\delta\tilde{\sigma}$ measured in our ensembles of glasses. Interestingly, in the IPL and HRTZ models we see that $\delta\tilde{\sigma}$ tends to decrease upon approaching the unjamming point by increasing $\beta$ (for IPL) or decreasing the pressure (for HRTZ), respectively. In contrast with our observations for e.g.~the Poisson's ratio showed in Fig.~\ref{poissons_ratio_fig}, no non-monotonic behavior in $\delta\tilde{\sigma}$ is observed in the IPL model. At $\beta\!\gtrsim\!16$ it appears that $\delta\tilde{\sigma}\!\sim\!\log \beta$.

%%%%%%%%%%%%%%%%%%%%%%%%%%%%%%%%%%%%%%%%%%%%%
\begin{figure*}[!ht]
\centering
\includegraphics[width = 1.00\textwidth]{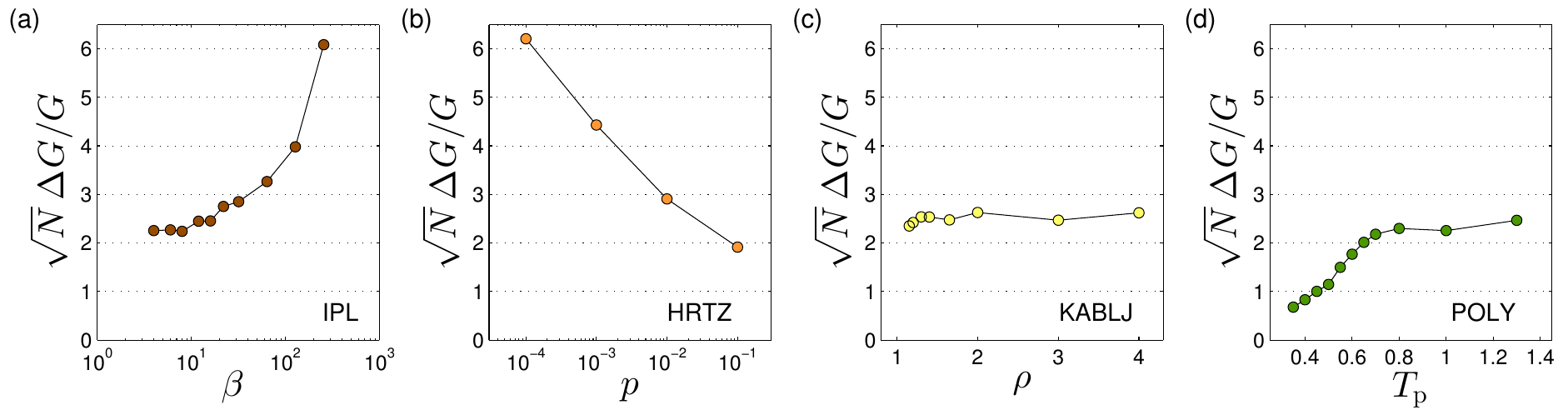}
\caption{\footnotesize $\Delta \tilde{G}\!\equiv\!\sqrt{N}\Delta G/G$ is a $N$-independent dimensionless quantifier of mechanical disorder, defined via Eqs.~(\ref{foo00}) and (\ref{foo01}), and motivated in Sect.~\ref{G_flucs_section}. The $p\!=\!10^{-5}$ data point for which $\Delta \tilde{G}\!=11.43$ was omitted for visual purposes. }
\label{G_flucs_fig}
\end{figure*}
%%%%%%%%%%%%%%%%%%%%%%%%%%%%%%%%%%%%%%%%%%%%%%%%%%%%%%

The KABLJ and POLY models appear to agree at high densities and high $T_{\rm p}$, respectively, showing $\delta\tilde{\sigma}\!\approx\!0.22$ in those regimes. The POLY system exhibits a significant reduction of $\delta\tilde{\sigma}$ upon annealing (i.e.~for lower $T_{\rm p}$), up to roughly 40\% below the high-$T_{\rm p}$ plateau value.

\subsection{Shear modulus fluctuations}
\label{G_flucs_section}
We next turn to characterizing the degree of mechanical disorder of our simple computer glasses. Following similar ideas put forward by Schirmacher and coworkers \cite{Maurer2004,Schirmacher_2006,Schirmacher_prl_2007}, we propose to quantify the mechanical disorder of a given ensemble of computer glasses by first measuring
\begin{equation}\label{foo00}
\Delta G \equiv \sqrt{\mbox{median}_i\big[ (G_i - G)^2\big]}\,,
\end{equation}
where the median is taken over the ensemble of glasses, and $G$ denotes the sample-to-sample mean shear modulus. In Appendix~\ref{N_scaling_appendix} we demonstrate that, as expected for an intensive variable (and see also \cite{exist}), $\Delta G\!\sim\!1/\sqrt{N}$. A dimensionless and $N$-independent quantifier of disorder is therefore given by 
\begin{equation}\label{foo01}
\Delta\tilde{G} \equiv \sqrt{N}\Delta G/G\,.
\end{equation}

In Fig.~\ref{G_flucs_fig} we plot $\Delta\tilde{G}$ for our different computer glasses. We find that $\Delta\tilde{G}$ grows substantially in the IPL and HRTZ models as the respective unjamming points are approached, suggesting that $\Delta\tilde{G}\!\to\!\infty$ upon approaching unjamming. 

While $\Delta G$ remains essentially constant at $\approx\!2.5$ over the entire density range in the KABLJ model, in the POLY model we find a very substantial decrease of $\Delta G$ as a function of the parent temperature $T_{\rm p}$, by over a factor of 3. The noise in our data is quite substantial; we nevertheless speculate based on our data that the variation rate $d (\delta \tilde{G})/dT_{\rm p}$ changes nonmonotonically with decreasing $T_{\rm p}$, namely that the decrease in $\delta\tilde{G}$ slows down at low $T_{\rm p}$. An interesting question to address in future studies is a possible relation between this nonmonotonicity with temperature, and that reported in \cite{Coslovich2018} for thermal activation barriers in deeply supercooled computer liquids. 

%%%%%%%%%%%%%%%%%%%%%%%%%%%%%%%%%%%%%%%%%%%%%
\begin{figure}[!ht]
\centering
\includegraphics[width = 0.49\textwidth]{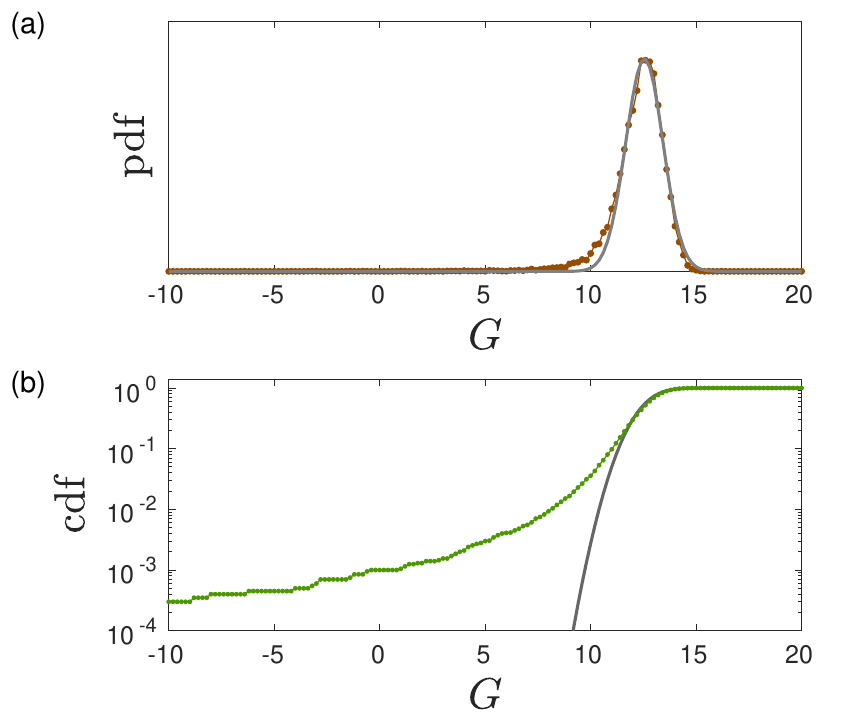}
\caption{\footnotesize (a) The dotted line represents the probability distribution function (pdf) $p(G)$ of the shear modulus of 20,000 computer glasses of $N\!=\!2000$ particles of the IPL model with $\beta\!=\!10$, made by a instantaneous quench from a high temperature liquid state. The continuous line is a fit to a Gaussian, that demonstrates the asymmetry of $p(G)$ about its mean. To better quantify the low-value tail of $p(G)$, in panel (b) we plot with a dotted line the cummulative distribution function (cdf) $\int^Gp(G')dG'$; the continuous line represents the cummulative distribution associated with the Gaussian fit of panel (a), shown for comparison. }
\label{G_dist_fig}
\end{figure}
%%%%%%%%%%%%%%%%%%%%%%%%%%%%%%%%%%%%%%%%%%%%%%%%%%%%%%

%%%%%%%%%%%%%%%%%%%%%%%%%%%%%%%%%%%%%%%%%%%%%
\begin{figure*}[!ht]
\centering
\includegraphics[width = 1.00\textwidth]{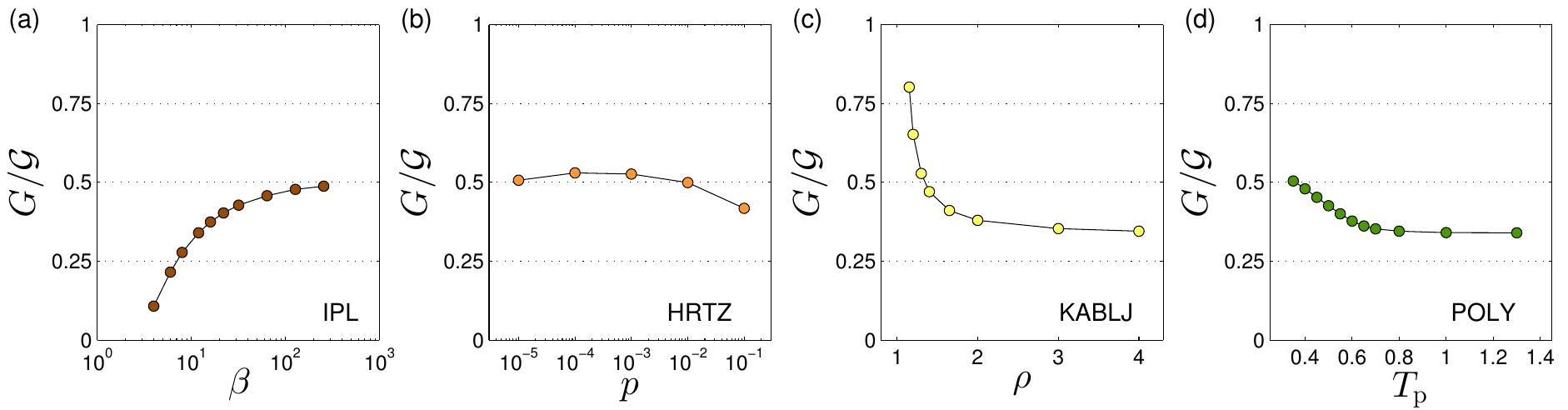}
\caption{\footnotesize The ratio $G/{\cal G}$ of the mean shear modulus to the mean unstressed shear modulus (see Sect.~\ref{observables} for definitions). Lower $G/{\cal G}$ indicates lower stability and an increasing role played by the interparticle forces in determining shear moduli.  }
\label{unstressed_G_fig}
\end{figure*}
%%%%%%%%%%%%%%%%%%%%%%%%%%%%%%%%%%%%%%%%%%%%%%%%%%%%%%

The reason we choose to measure the median of fluctuations instead of the considering the more conventional standard deviation is that for small $N$ the distribution $p(G)$ of the shear modulus can feature a large tail at low values. This is demonstrated in Fig.~\ref{G_dist_fig}, where we show the distribution of shear moduli measured in the IPL model for glasses of $N\!=\!2000$ particles that were instantaneously quenched from high temperature states. Fig.~\ref{G_dist_fig}b shows that the low-$G$ tail is substantial, leading to a large discrepancy between the full width at half maximum of the distribution of $G$ and its standard deviation. To overcome this discrepancy we opt for a measure which is based on the (square root of the) median of fluctuations rather than their mean. We note however that the large tail of $p(G)$ at low values of $G$ is expected to disappear as the system size is increased \cite{exist}.

\subsection{Effect of internal stresses on shear modulus}

We conclude our study of the elastic properties of our computer glasses with presenting and discussion the effect of internal stresses on the shear modulus. To this aim we recall Eq.~\ref{mock_U} which defines a modified potential energy ${\cal U}$, constructed based on the original potential energy $U$ by connecting a relaxed, Hookean spring between all pairs of interacting particles, with stiffnesses $k\!=\!\varphi''|_{r_{ij}}$ adopted from the original pairwise potentials $\varphi$. An associated shear modulus ${\cal G}$ is then defined as $V^{-1}\frac{d^2{\cal U}}{d\gamma^2}$. In previous work \cite{eric_boson_peak_emt} it has been shown using mean field calculations that $G/{\cal G}$ indicates the distance of a system from an internal-stress-induced elastic instability. It is predicted in \cite{eric_boson_peak_emt} that $G/{\cal G}\!\approx\!1/2$ in marginally-stable states with harmonic pairwise interactions, and $G/{\cal G}\!>\!1/2$ as glass stability increases. The ratio $G/{\cal G}$ can also depend on statistical properties of interparticle interactions, as discussed in \cite{eric_hard_spheres_emt}. 

In Fig.~\ref{unstressed_G_fig} we show measurements of $G/{\cal G}$ in our different computer glasses. In the IPL model we find that $G/{\cal G}\!<\!1/2$ in the entire $\beta$ range, but approaches $1/2$ in the large $\beta$ limit at which the system unjams. In the HRTZ system we find $G/{\cal G}\!\approx\!1/2$ over most of the investigated pressure range, with a slight decrease at high pressures. The KABLJ system shows that $G/{\cal G}$ can attain high values in the low density regime in which attractive interactions become dominant, and, similarly to as we have seen above, at large densities it agrees well with the $\beta\!\approx\!12$ result for $G/{\cal G}$ of the IPL model. Finally, in the POLY system at high $T_{\rm p}$, $G/{\cal G}$ agrees well with the IPL model for $\beta\!\approx\!12$, as expected. Equilibration deep into the supercooled regime increases $G/{\cal G}$ by nearly 50\%, bringing it to $\approx\!1/2$ at the deepest supercooling. 

%%%%%%%%%%%%%%%%%%%%%%%%%%%%%%%%%%%%%%%%%%%%%
\begin{figure*}[!ht]
\centering
\includegraphics[width = 0.75\textwidth]{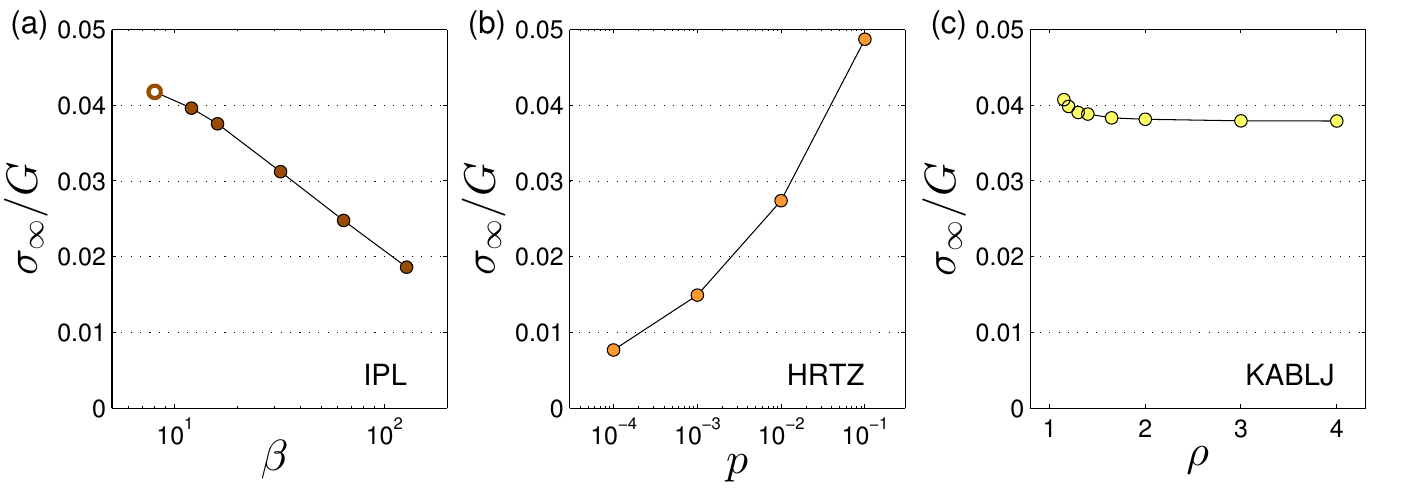}
\caption{\footnotesize Yield stress $\sigma_\infty$ rescaled by the mean isotropic, as-cast shear modulus $G$. We reiterate that the open symbol in panel (a) represents an approximation obtained using the finite-cutoff variant of the IPL pairwise potential, see Sect.~\ref{ipl_model} for details. We further note that data points for $p\!=\!10^{-5}$ in the HRTZ model and $\beta\!=\!256$ in the IPL model could not be measured due to numerical convergence difficulties.}
\label{yield_stress_figure}
\end{figure*}
%%%%%%%%%%%%%%%%%%%%%%%%%%%%%%%%%%%%%%%%%%%%%%%%%%%%%%

\subsection{Yield stress}

Up untill this point we have only discussed various dimensionless characterizations of the elastic properties of our computer glasses. In this last Subsection we present results regarding the simple shear yield stress of a subset of the models we have investigated, measured in athermal quasistatic plastic flow simulations. In particular, we exclude the POLY model from this analysis; its elasto-plastic transient behavior was characterized in detail in \cite{Ozawa6656}, and its steady-flow state stress (referred to here as the yield stress) is expected to be independent of the key control parameter of the POLY model -- the parent temperature $T_{\rm p}$. 

We employ the standard procedure for driving our glasses under athermal quasistatic deformation: the simulations consist of repeatedly applying a simple shear deformation transformation (we use strain steps of $\Delta\gamma\!=\!10^{-3}$), followed by a potential energy minimization under Lees-Edwards boundary conditions \cite{allen1989computer}. As explained in Sect.~\ref{HRTZ_model}, simulations of the HRTZ model involved embedding a barostat functionality \cite{berendsen} into our minimization algorithm, in order to maintain the pressure approximately constant during the deformation simulations, see further discussion in Appendix~\ref{barostat_hrtz}.

In Fig.~\ref{yield_stress_figure} we present the average yield stress $\sigma_\infty$, defined here as the average steady-flow stress, taken after the initial elastoplastic transients, rescaled by the isotropic-states average shear modulus $G$. Each point is obtained by averaging over the steady flow shear stress of 200 independent runs of each computer glass model, and for each control parameter value.

We find that in the IPL and HRTZ models $\sigma_\infty/G$ decreases upon approaching their respective unjamming points $\beta\!\to\!\infty$ and $p\!\to\!0$. In the IPL model we observe $\sigma_\infty/G\!\sim\!\log\beta$ at large $\beta$; understanding this behavior is left for future investigations. In the HRTZ model one expects $\sigma_\infty\!\sim\! p$ and $G\!\sim\! p^{1/3}$ (it should scale with pressure similarly to the bulk modulus $K$ of isotropic, as-cast states, see \cite{Baity-Jesi2017}), then $\sigma_\infty/G \!\sim\! p^{2/3}$ is predicted. We cannot however confirm this prediction numerically; we postulate that the pressure range explored is not sufficiently close to the unjamming point in order to observe the asymptotic scaling. Finally, the KABLJ model features $\sigma_\infty/G\!\approx\!0.038$ over the majority of the explored density range, with a slight increase as attractive forces become more dominant at low densities. 

How do these numbers compare to more realistic computer glasses? In \cite{CHENG20093253} values of around $\sigma_\infty/G\!\approx\!0.05$ were reported for a model $\mbox{Cu}_{64}\mbox{Zr}_{36}$ metallic glasses that employs the embedded atom method \cite{PhysRevLett.50.1285}, i.e.~some 30\% higher than what we find in e.g.~the KABLJ model. Similar results were also found by \cite{Wang2018} for model $\mbox{Cu}_{50}\mbox{Zr}_{50}$ and $\mbox{Cu}_{47.5}\mbox{Zr}_{47.5}\mbox{Al}_5$ metallic glasses. In \cite{Argon_prb_2005} a value of $\sigma_\infty/G\!\approx\!0.03$ was observed using the Stillinger-Weber model for amorphous silicon \cite{Stillinger_Weber}. A value of $\sigma_\infty/G\!\approx\!0.11$ can be estimated based on the stress-strain signals reported in \cite{MOLNAR2016129} for computer models of sodium silicate glasses that employ the van Beest-Kramer-van Santen potential \cite{PhysRevLett.64.1955}. The spread in these values indicates that the simple computer models investigated in this work only represent a narrow class of amorphous solids.

%In \cite{Samwer_prl_2005} measurements of the shear stress at yielding $\sigma_Y$ (which can differ from the yield stress $\sigma_\infty$ studied here) are presented for a large number of metallic glasses. It is found that $\sigma_Y/G\!\approx0.053$ for several metallic glasses, which is quite significantly higher than the values of $\sigma_\infty/G$ measured in our generic computer glasses such as the IPL (for intermediate $\beta$) and the KABLJ. However, the shear stress at yielding can be substantially larger than the steady flow stress (see e.g.~\cite{Harmon_apl_2007}), explaining the disagreement with our data. 

\section{Summary}
\label{summary}
The goal of this paper is to offer a comprehensive data set that compares --- on the same footing --- various dimensionless quantifiers of elastic and elasto-plastic properties of popular computer glass models. We build on the assertion that instantaneously quenching high-energy configurations to zero temperature defines an ensemble of glassy samples that can be meaningfully compared between different models. We aimed at disentangling the effects on mechanical properties of various features of the interaction potentials that define computer glass models, from those induced by varying external control parameters and preparation protocols.  We hope that the various data sets presented in this work, and the dimensionless observables put forward in this work, will be used as a benchmark for future studies, allowing to meaningfully compare the mechanical properties of different computer glass models.

In addition to putting forward our various analyses of mechanical properties of computer glasses, we have also made a few new observations, summarized briefly here: we have identified an interesting nonmonotonicity in the Poisson's ratio in the IPL model (see Fig.~\ref{poissons_ratio_fig}), as a function of the exponent $\beta$ of the inverse-power-law interactions. The shear-to-bulk moduli ratio $G/K$ echos this nonmonotonicity: $G/K$ decreases dramatically as $\beta$ is made small, in addition to its expected decrease at large $\beta$ -- the limit at which the IPL model experiences an unjamming transition \cite{stefanz_pre}. We have shown that the small-$\beta$ decrease is due to the increasingly dominant role of the pressure in determining the shear modulus, in parallel to the decreasing role of the nonaffine, relaxation term.

Importantly, we have shown that the KABLJ model features a Poisson's ratio that resembles that of laboratory metallic glasses, and, at density of order unity is generally lower than that seen for the purely repulsive and isomorph-invariant \cite{Dyre_2016} IPL model; our study indicates that the increased nonaffinity of the bulk modulus at low pressures plays an important role in determining the Poisson's ratio in the KABLJ model. 

We offered a dimensionless quantifier of internal glassy frustration, $\delta\tilde{\sigma}$, shown to decrease by up to $40\%$ in well annealed glasses compared to poorly annealed glasses. Even more remarkable is the annealing-induced variation in the sample-to-sample relative fluctuations of the shear modulus $\Delta\tilde{G}$ (cf.~Eq.~(\ref{foo01}) and Fig.~\ref{G_flucs_fig}d), that decrease by over a factor of 3 between poorly annealed and well annealed glasses. Finally, an intriguing nonmonotonic behavior of $d (\delta \tilde{G})/dT_{\rm p}$ with equilibrium parent temperature $T_{\rm p}$ was also observed. 

An observable inaccesible experimentally but easily measured numerically is the ratio $G/{\cal G}$ of the shear modulus $G$ to that obtained by removing the internal forces between particles, denoted here and above by ${\cal G}$. A similar procedure was carried out in previous work in the context of the vibrational spectrum of glasses \cite{eric_boson_peak_emt,inst_note,ikeda_pnas}, and for the investigation of the lengthscale associated with the unjamming point \cite{breakdown}. In theoretical work \cite{eric_boson_peak_emt,eric_hard_spheres_emt} some trends are predicted for $G/{\cal G}$; however, since it varies both with stability and depends on details of the interaction potential, it usefulness as a characterizer of stability of a computer glass appears to be limited.

\acknowledgements
We warmly thank Eran Bouchbinder, Geert Kapteijns, David Richard and Eric DeGiuli for discussions and for their useful comments on the manuscript. Support from the Netherlands Organisation for Scientific Research (NWO) (Vidi grant no.~680-47-554/3259) is acknowledged.

\appendix

\section{Cutoff and finite-size effects in the IPL model}
\label{ipl_appendix}

In this Appendix we show the effects of the dimensionless cutoff $x_c$ of the pairwise potential (see Sect.~\ref{ipl_model}) and of the system size $N$ on the bulk modulus $K$, and motivate our choices of system sizes and cutoffs used in our main analyses. We first note that in the full-blown $N\!\to\!\infty$ IPL model the nonaffine term of the bulk modulus (see Eqs.~(\ref{bulk_modulus_def}) and (\ref{nonaffine_term})) is identically zero. In finite, periodic IPL systems with a finite cutoff in the pairwise potential, the nonaffine term is not identically zero, but still negligibly small. Next, we see that neglecting the nonaffine term term of the bulk modulus, for the case of pairwise potentials one has
\begin{equation}
K \simeq \frac{1}{V}\sum_{i<j}\varphi_{ij}'' r_{ij}^2\,.
\end{equation}
If the pairwise interaction is cutted-off at $\lambda x_c$ ($\lambda$ is a microscopic length), then the bulk modulus can be estimated as
\begin{eqnarray}
K(x_c) \sim \int_0^{\lambda x_c}r^2 \varphi'' r^2 dr \sim K_\infty - \int_{\lambda x_c}^\infty r^{2-\beta}dr \,,
\end{eqnarray}
and therefore the deviation from the $x_c\!\to\!\infty$ value should follow
\begin{equation}\label{cutoff_effect}
K_\infty - K(x_c) \sim x_c^{3-\beta}\,.
\end{equation}
A similar consideration would also apply to the effect of system size, setting $x_c\!\approx\! L/2$, namely
\begin{equation}\label{N_effect}
K_\infty - K(N) \sim N^{\frac{3-\beta}{3}}\,.
\end{equation}

%%%%%%%%%%%%%%%%%%%%%%%%%%%%%%%%%%%%%%%%%%%%%
\begin{figure}[!ht]
\centering
\includegraphics[width = 0.50\textwidth]{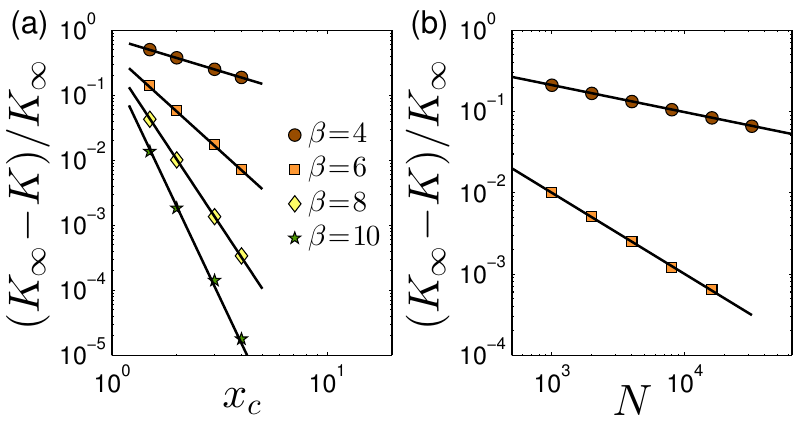}
\caption{\footnotesize (a) Effect of dimensionless cutoff $x_c$ on the bulk modulus, for systems of $N\!=\!8000$ particles. The relative deviation below an asymptotic $K_\infty$ is given by Eq.~(\ref{cutoff_effect}), i.e.~$\sim x_c^{3-\beta}$ represented by the continuous lines. (b) Effect of system size on the bulk modulus, for $x_c\!\approx\! L/2$. For $\beta\!\ge\!8$ (not shown) the sample-to-sample noise is larger than the deviation from the asymptotic $K_\infty$, which are the same as used for panel (a). The continuous lines follow $N^{(3-\beta)/3}$, as given by Eq.~(\ref{N_effect}).}
\label{N_dependence_ipl}
\end{figure}
%%%%%%%%%%%%%%%%%%%%%%%%%%%%%%%%%%%%%%%%%%%%%%%%%%%%%%

In Fig.~\ref{N_dependence_ipl} we show data validating Eqs.~(\ref{cutoff_effect}) and (\ref{N_effect}). From these data we conclude that the finite size effect on the bulk modulus are smaller than 0.1\% for $\beta\!\ge\!6$ and $N\!\ge\!8000$ (when the the cutoff $x_c\!\approx\!L/2$), and that the effect of a finite cutoff is smaller than 1\% for $x_c\!=\!1.5$ and $\beta\!>\!10$. For these reasons, we employ the long-range cutoff $x_c\!\approx\! L/2$ for $\beta\!<16$, and employ systems of $N\!=\!32000$ and $N\!=\!16000$ for $\beta\!=\!4$ and $\beta\!=\!6$, respectively, and $N\!=\!8000$ otherwise. We employ the short range cutoff $x_c\!=\!1.5$ for $\beta\!\ge\!16$.

%%%%%%%%%%%%%%%%%%%%%%%%%%%%%%%%%%%%%%%%%%%%%
\begin{figure}[!ht]
\centering
\includegraphics[width = 0.50\textwidth]{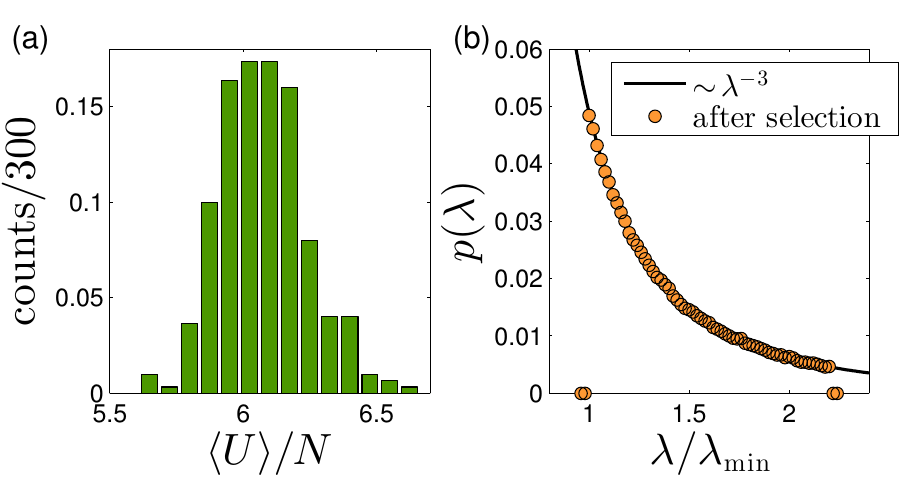}
\caption{\footnotesize (a) Distribution of mean energy per particle, measured for 1000 independent realizations equlibrated at $T\!=\!0.60$. (b) Distribution of post-selection particle size parameters, see text for details.}
\label{handle_fluctuations}
\end{figure}
%%%%%%%%%%%%%%%%%%%%%%%%%%%%%%%%%%%%%%%%%%%%%%%%%%%%%%

\section{Sample-to-sample realization fluctuations}
\label{fluctuations}
The POLY model employed in this work considers soft spheres with polydispersed size parameters, which are drawn from a distribution $p(\lambda)\!\sim\!\lambda^{-3}$ sampled between $\lambda_{\mbox{\tiny min}}$ and $\lambda_{\mbox{\tiny max}}$ \cite{berthier_prx}, see Sect.~\ref{swap}. Following \cite{berthier_prx}, we chose $\lambda_{\mbox{\tiny max}}/\lambda_{\mbox{\tiny min}}\!=\!2.22$; this choice can lead to large fluctuations between the energetic and elastic properties of different finite-size samples. To demonstrate this, we show in Fig.~\ref{handle_fluctuations}a the distribution of the mean energies (per particle), calculated over 1000 independent equilibrium runs at $T\!=\!0.6$, each run pertaining to a different, independent realization of the particle-size parameters $\lambda_i$ drawn from the same parent distribution $p(\lambda)$, and with $N\!=\!8000$ particles. The mean (over realizations) standard deviation of the energy per particle of individual runs was found to be $\approx\!0.01$, whereas the standard deviation (over realizations) of the mean energy per particle is $\approx\!0.16$, i.e.~much larger than the characteristic energy per particle fluctuations of any given realization of particle-size parameters, for $N\!=\!8000$. 

%%%%%%%%%%%%%%%%%%%%%%%%%%%%%%%%%%%%%%%%%%%%%
\begin{figure}[!ht]
\centering
%\vspace{0.5cm}
\includegraphics[width = 0.50\textwidth]{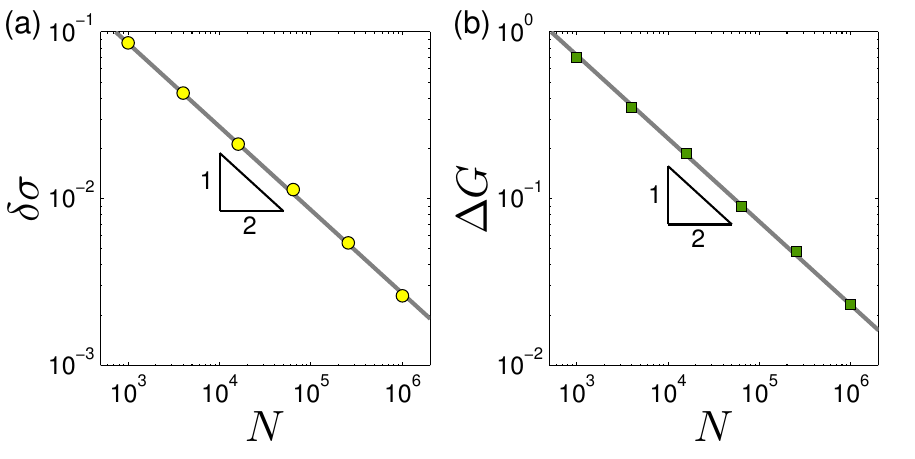}
\caption{\footnotesize (a) Sample-to-sample standard deviations $\delta\sigma$ of the as-cast shear stress, plotted against system size $N$. (b) The measure $\Delta G$ (cf.~Eq.~(\ref{foo00})) vs.~system size $N$. Both measures of fluctuations decay as $1/\sqrt{N}$.}
\label{stress_and_G_flucs_N_scaling}
\end{figure}
%%%%%%%%%%%%%%%%%%%%%%%%%%%%%%%%%%%%%%%%%%%%%%%%%%%%%%

In order to minimize the effects of these finite-size fluctuations, we selected the particular realizations whose mean equilibrium energy deviated from the mean \emph{over realizations} (measured here to be $\approx\!6.114$) by less than 0.5\%, and discard of the rest. To test whether this selection protocol has any observable effect on the distribution of particle size parameters, in Fig.~\ref{handle_fluctuations}b we plot the distribution of particle size parameters measured only in the selected states. We find no observable effect of discarding of the realizations with too large or too small energies --- as described above --- on the distribution of particle size parameters.

\section{System size scaling of fluctuations}
\label{N_scaling_appendix}
In Sections~\ref{internal_stress} and~\ref{G_flucs_section} we define two dimensionless measures of elastic properties of glasses: $\delta\tilde{\sigma}\!\equiv\!\sqrt{N}\delta\sigma/G$ and $\Delta\tilde{G}\!\equiv\!\sqrt{N}\Delta G/G$, respectively, where $\delta\sigma$ denotes the standard deviation of the as-cast shear stress $\sigma$, and $\Delta G$ is a measure of fluctuations that follows the definition given by Eq.~\ref{foo00}. To establish that $\delta\tilde{\sigma}$ and $\Delta\tilde{G}$ are independent of system size $N$, in Fig.~\ref{stress_and_G_flucs_N_scaling}a we plot $\delta\sigma$ vs.~system size $N$, and in Fig.~\ref{stress_and_G_flucs_N_scaling}b we plot $\Delta G$ vs.~$N$. The model glass employed is the IPL model with $\beta\!=\!10$ \cite{cge_paper}. As asserted, both of these observables depend on system size as $1/\sqrt{N}$, implying the $N$-independence of $\delta\tilde{\sigma}$ and~$\Delta\tilde{G}$. 

%%%%%%%%%%%%%%%%%%%%%%%%%%%%%%%%%%%%%%%%%%%%%
\begin{figure}[!ht]
\centering
\vspace{0.5cm}
\includegraphics[width = 0.50\textwidth]{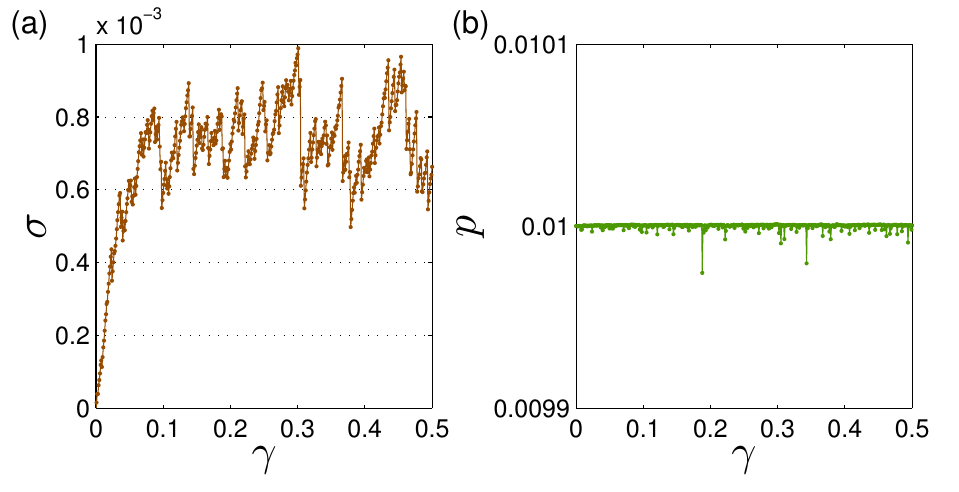}
\caption{\footnotesize (a) Stress vs.~strain measured in a quasistatic shear deformation simulation of the HRTZ model at constant external pressure of $p\!=\!10^{-2}$. (b) Pressure vs.~strain in the same run shown in (a). Small fluctuations of less than 1\% are still observed; our numerical scheme does not fix the pressure exactly, but rather only approximately.}
\label{constant_p_deformation_fig}
\end{figure}
%%%%%%%%%%%%%%%%%%%%%%%%%%%%%%%%%%%%%%%%%%%%%%%%%%%%%%

\section{Athermal quasistatic simulations of the HRTZ model at fixed external pressure}
\label{barostat_hrtz}
The key control parameter of the HRTZ model is the external pressure $p$; when creating glassy samples of this model, we incorporated a numerical scheme \cite{berendsen} that allows to specify the desired target pressure into our potential energy minimization algorithm. While this scheme does not fix the pressure exactly, it is sufficiently accurate for our purposes. The performance of the fixed pressure protocol in our quasistatic shear simulations can be gleaned from the example signals shown in Fig.~\ref{constant_p_deformation_fig}.

%\bibliography{references_lerner}

\end{document}